\documentclass[aps,prl,twocolumn,superscriptaddress,10pt,english,preprintnumbers]{revtex4-1}
\pdfoutput=1
\usepackage[T1]{fontenc}
\usepackage[utf8]{inputenc}
\usepackage{epsfig}
\usepackage{graphicx}
\usepackage{xcolor}
\usepackage{latexsym}
\usepackage{textcomp}
\usepackage{amssymb}
\usepackage[colorlinks=true,linkcolor=blue,citecolor=blue, urlcolor=blue]{hyperref} 
\usepackage{amsfonts,amsthm,amstext,amscd}
\usepackage{amsmath}
\usepackage{mathtools}

\def\trento{T\raisebox{-0.5ex}{R}ENTo}

\def\tr{\emph{Trajectum}}

\def\saa{\sigma_{AA}}
\def\snn{\sigma_{NN}}
\def\trento{T\raisebox{-0.5ex}{R}ENTo}

\begin{document}
\title{The hadronic nucleus-nucleus cross section and the nucleon size}
\author{Govert Nijs}
\affiliation{Center for Theoretical Physics, Massachusetts Institute of Technology, Cambridge, MA 02139, USA}
\author{Wilke van der Schee}
\affiliation{Theoretical Physics Department, CERN, CH-1211 Gen\`eve 23, Switzerland}
\begin{abstract}
Even though the total hadronic nucleus-nucleus cross section is among the most fundamental observables, it has only recently been measured precisely for lead-lead collisions at the LHC\@. This measurement implies the nucleon width should be below 0.7 fm, which is in contradiction with all known state-of-the-art Bayesian estimates. 
We study the implications of the smaller nucleon width on quark-gluon plasma properties such as the bulk viscosity. The smaller nucleon width dramatically improves the description of several triple-differential observables.
\end{abstract}

\preprint{CERN-TH-2022-103//MIT-CTP/5445}

\maketitle

{\bf Introduction -} The understanding of the creation quark-gluon plasma (QGP) as created at colliders such as the Large Hadron Collider (LHC) in Geneva requires the understanding of several stages of the collision of heavy ions \cite{1301.2826, 1802.04801}\@. The first stage is far-from-equilibrium and involves an initial condition together with an understanding how this evolves towards a hydrodynamic QGP \cite{2005.12299}. Secondly, there is a hydrodynamic stage, which involves understanding the temperature-dependent transport coefficients such as the shear viscosity \cite{1712.05815}\@. Lastly, the QGP undergoes particlization into a gas of interacting hadrons which can then be detected experimentally.

Relatively little is known about the initial stage and many analyses use a phenomenological parameterization called the \trento{} model \cite{1412.4708}\@. Here a heavy ion is composed of a superposition of nucleons with Gaussian distributions of energy of width $w$, while fixing the interaction rate to fit the hadronic nucleon-nucleon cross section $\snn$ determined from $pp$ collisions. Hence, interactions of nucleons with smaller $w$ depend strongly on the impact parameter, while large $w$ nucleons do not always interact even if they collide at zero impact parameter. Interestingly, all recent state-of-the-art global analyses of a wide variety of experimental data have preferred a large nucleon width in fm of $0.98 \pm 0.18$ \cite{1808.02106}, $0.96\pm 0.05$ \cite{Bernhard:2019bmu}, $0.94\pm0.18$ \cite{2010.15130}, $1.05\pm 0.13$ \cite{2011.01430} or $0.82 \pm 0.23$ \cite{2110.13153}\@. 
Since $w$ equals the Gaussian width this implies that the resulting energy profile is then much larger than the proton charge radius of $0.841\,$fm \cite{Pohl:2010zza}, or the radius implied by $\snn = 68\,$mb in the black disk approximation ($r=0.74\,$fm).

In this Letter we show that the recent ALICE measurement of the PbPb total hadronic cross section $\saa{}$ of $7.67\pm 0.24\,$b at $\sqrt{s_{\rm NN}} = 5.02\,$TeV \cite{2204.10148manual} implies a nucleon width smaller than approximately $0.7\,$fm, which is smaller than the width from all quoted Bayesian estimates. This measurement hence raises two important questions. First, why did the Bayesian probability estimates not result in the correct nucleon width? Second, what are the implications of this smaller nucleon width? Part of the answer to the first question must be an inaccurate estimate of the systematic uncertainty covariance matrix. Here we note that the covariance matrix does not only include the systematic and statistical experimental and theoretical uncertainties, but it is essential to also include correlations or anti-correlations between observables. For the second question, the smaller nucleon width implies a larger bulk viscosity. Finally, we will show the improved analysis implies a better description of statistically difficult triple-differential observables.

{\bf The initial condition and the cross section - } All computations of  $\saa$ start with the Monte Carlo Glauber model \cite{0805.4411, 1408.2549, 1710.07098, 2011.14909}. In this model nucleons are placed randomly according to a Woods-Saxon distribution, $\rho(r)=\rho_0/(1+e^{(r-R)/a})$, specified by the half-density radius $R$ and the diffusivity $a$ with potentially the requirement of a minimum nucleon-nucleon distance $d_{\rm min}$. In the current implementation the parameters $R$ and $a$ follow the point density distributions that are determined separately for protons and neutrons using electron-ion scattering experiments \cite{Klos:2007is, 1311.0168, 1710.07098}\@. $\saa$ is then determined by the condition that at least a single nucleon-nucleon interaction occurs.

In the black disk approximation nucleons interact if their distance $d$ satisfies $d<\sqrt{\snn/\pi}$, with $\snn$ the nucleon-nucleon cross section as determined from $pp$ collisions at the same collision energy.
In Pythia 8 a normalized overlap function is specified as a function of the impact parameter $b$ as $T_{pp}(b) \propto \exp(-(b/w)^m)$ with $w$ the nucleon width and $m=1.85$ for the Monash tune \cite{1404.5630,2011.14909}\@. In this work we will use the \trento{} model \cite{1412.4708}, which uses a Gaussian overlap function ($m=2$) and the nucleon width $w$ as a parameter. Nucleons then interact with probability $P(b)=1-\exp (-\sigma_{gg}\int d^2x_T T_A(x_T)T_B(x_T))$, whereby $\sigma_{gg}$ is determined by $\snn{}$\@.
In this work we use nucleons that are composed of smaller constituents (as in \cite{1808.02106}); we verified that this does not affect $\saa$\@.

It is no surprise that the (traditionally used) black-disk approximation produces the smallest $\saa$\@. Indeed, all models by construction have an equal nucleon-nucleon cross section, but wider overlap functions allow a nucleon-nucleon interaction to occur more easily for a nucleus-nucleus collision at relatively large impact parameter.

\begin{figure}[h]
\includegraphics[width=0.42\textwidth]{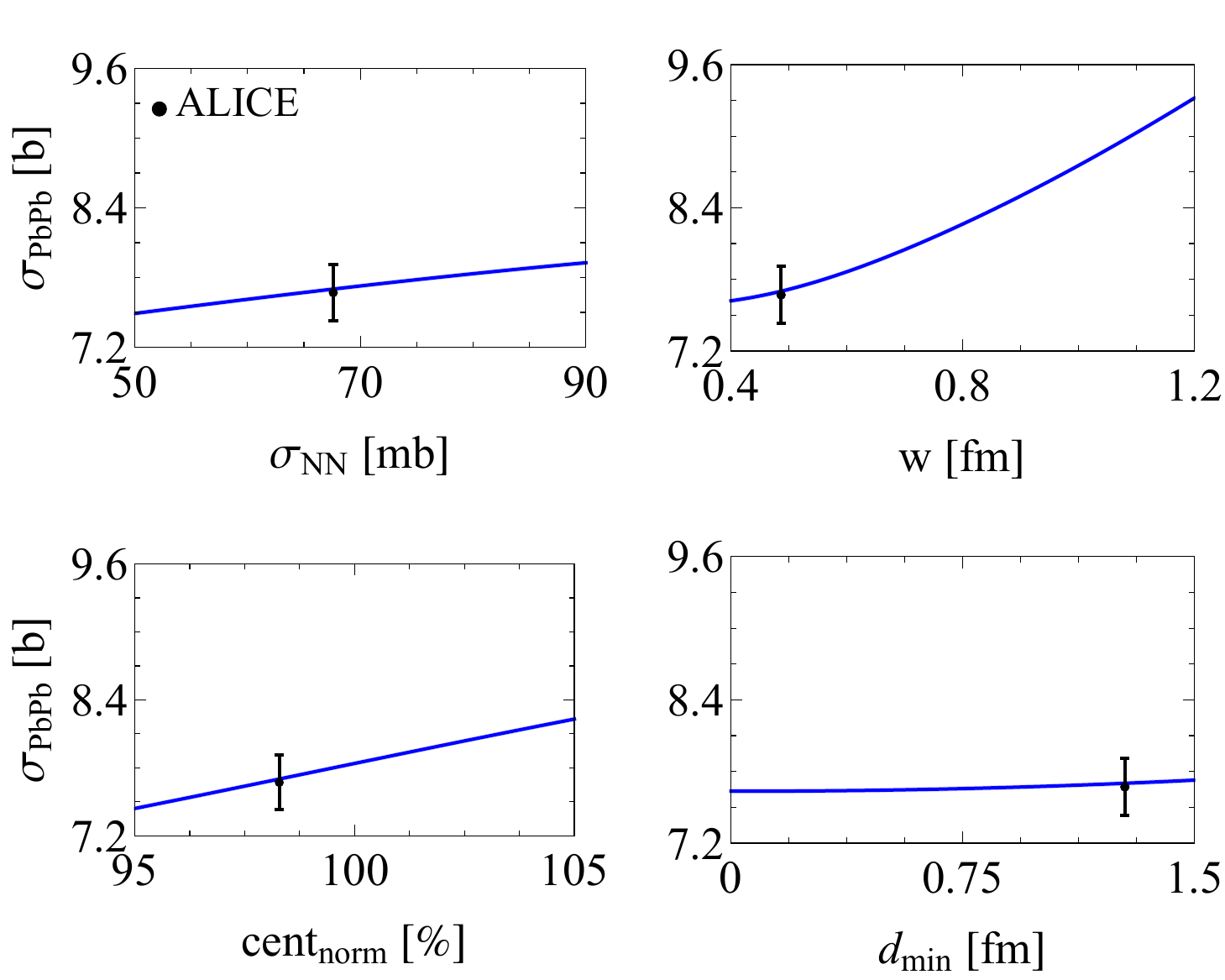}
\caption{\label{fig:crosssectionemulator}We show the dependence of the PbPb cross section on the nucleon-nucleon cross section $\sigma_\text{NN}$, the nucleon width $w$, the centrality normalization $\text{cent}_\text{norm}$ and the minimal nucleon-nucleon distance $d_\text{min}$\@. The parameters other than the one varied are kept fixed at the value indicated by the ALICE data point \cite{2204.10148manual}.}
\end{figure}

In Fig.~\ref{fig:crosssectionemulator} we see that $\saa$ can increase up to 23\% for $w$ as large as $1.2\,$fm. Perhaps surprisingly, the dependence on $\snn$ is fairly mild, and in this work we will keep the measured value of $61.2$ and $67.6\,$mb for 2.76 and 5.02 TeV collisions respectively \cite{1710.07098}\@. We also observe an almost negligible dependence on $d_{\rm min}$\@. 
The cross section depends linearly on the centrality normalization \cite{2110.13153}, e.g.~on which events to count as a collision (as in the models above), or experimentally on how many collisions are recorded. Both theoretically and experimentally this contains an uncertainty, which we include as a separate parameter in Fig.~\ref{fig:crosssectionemulator} (bottom-left)\@. Motivated by \cite{2204.10148manual} we give this parameter a prior probability distribution for our Bayesian analysis of a Gaussian with unit width. We note that this parameter propagates into a significant uncertainty mainly for peripheral spectra and multiplicities, but also for more central elliptic flow coefficients.

Crucially, $\saa$ only depends sensitively on $w$, $\snn$ and the centrality normalization, whereby the latter two are well constrained experimentally. A measurement of $\saa$ hence provides robust constraints on $w$ with only weak theoretical modelling uncertainties. The recently measured value of $7.67\pm 0.26\,$b implies $w\approx \text{0.4 -- 0.5}\,$fm (see Fig.~\ref{fig:crosssectionemulator}), which as noted in the introduction is in direct contradiction with all state-of-the-art global analyses of heavy ion collisions so far.

\begin{figure}[ht]
\includegraphics[width=0.46\textwidth]{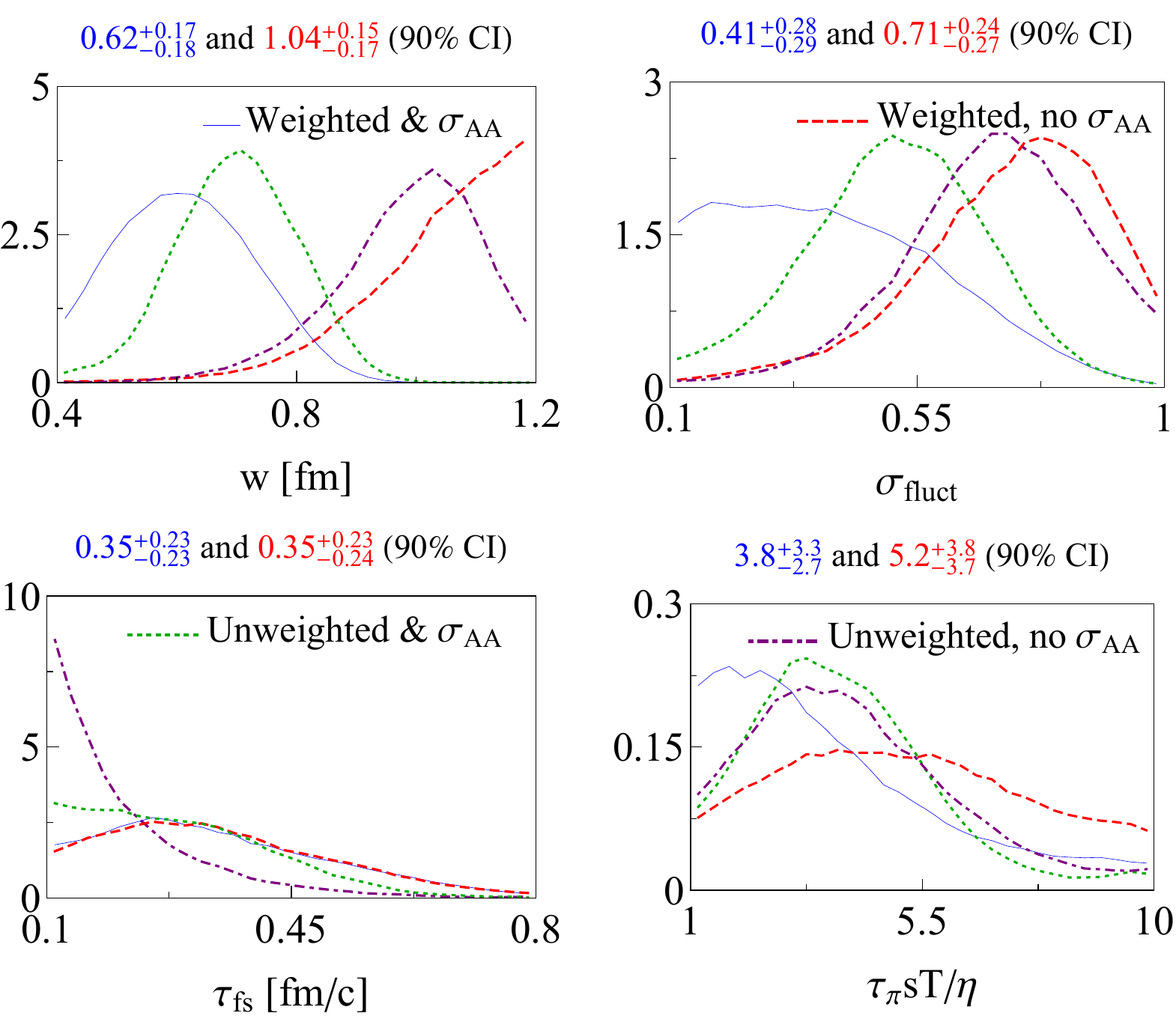}
\caption{\label{fig:postparams}We show posterior distributions for the nucleon width $w$, initial state fluctuation parameter $\sigma_\text{fluct}$, free streaming time $\tau_\text{fs}$ and shear relaxation time $\tau_\pi$\@. We show the results from four different fits, where we turn weighting on and off, and do or do not fit to the cross section $\sigma_\text{AA}$\@.}
\end{figure}

{\bf Implications for QGP properties - }
Fig.~\ref{fig:postparams} shows the posterior parameter analysis for the width $w$, the \trento{} fluctuation parameter $\sigma_{\rm fluct}$, the free streaming time $\tau_\text{fs}$ and the second order hydro relaxation time $\tau_\pi$ when including (green dotted) or excluding (purple dot-dashed) $\saa$\@. This is the result of a global analysis of 653 experimental data points \cite{1012.1657, 1512.06104,1603.04775,1805.05212,1910.07678,1303.0737,1407.5530,1804.02944,1606.06057,1805.04390,2204.10148manual,1509.03893} (see also the Supplemental Material) with a 21-dimensional parameter space within the publicly available \emph{Trajectum} 1.3 framework \cite{trajectumcode} similar to \cite{2010.15130,2010.15134,2110.13153}, but full details will be presented elsewhere \cite{companion}\@. The analysis with $\saa{}$ also includes the proton-lead cross section collisions at $\sqrt{s_{\rm NN}} = 5.02\,$TeV of $\sigma_{p\text{Pb}} = 2.06\pm 0.08\,$b as measured earlier by CMS \cite{1509.03893}\@. It is striking that adding two experimental observations ($\saa$ for PbPb \cite{2204.10148manual} and $p$Pb at 5.02 TeV \cite{1405.1849, 1509.03893}) has such a strong effect on these parameters.

Nevertheless, we see that the nucleon width is still not quite compatible with the value required from the $\saa$ measurement (the green dotted $w$ is peaked around $0.7\,$fm). This is consistent with the previous Bayesian analyses that strongly prefer larger widths and hence a Bayesian updated estimate now reduces the width, but not all the way to be compatible. Given the theoretical robustness of the $\saa$ comparison this remaining discrepancy has to be taken seriously. A crude way would be to fix $w=0.45\,$fm by hand or to use a prior distribution close to this value. A more quantitative approach is to give higher weights to observables we have a higher trust in on physical grounds, which is what we will attempt in this work. 

\begin{figure}[ht]
\includegraphics[width=0.46\textwidth]{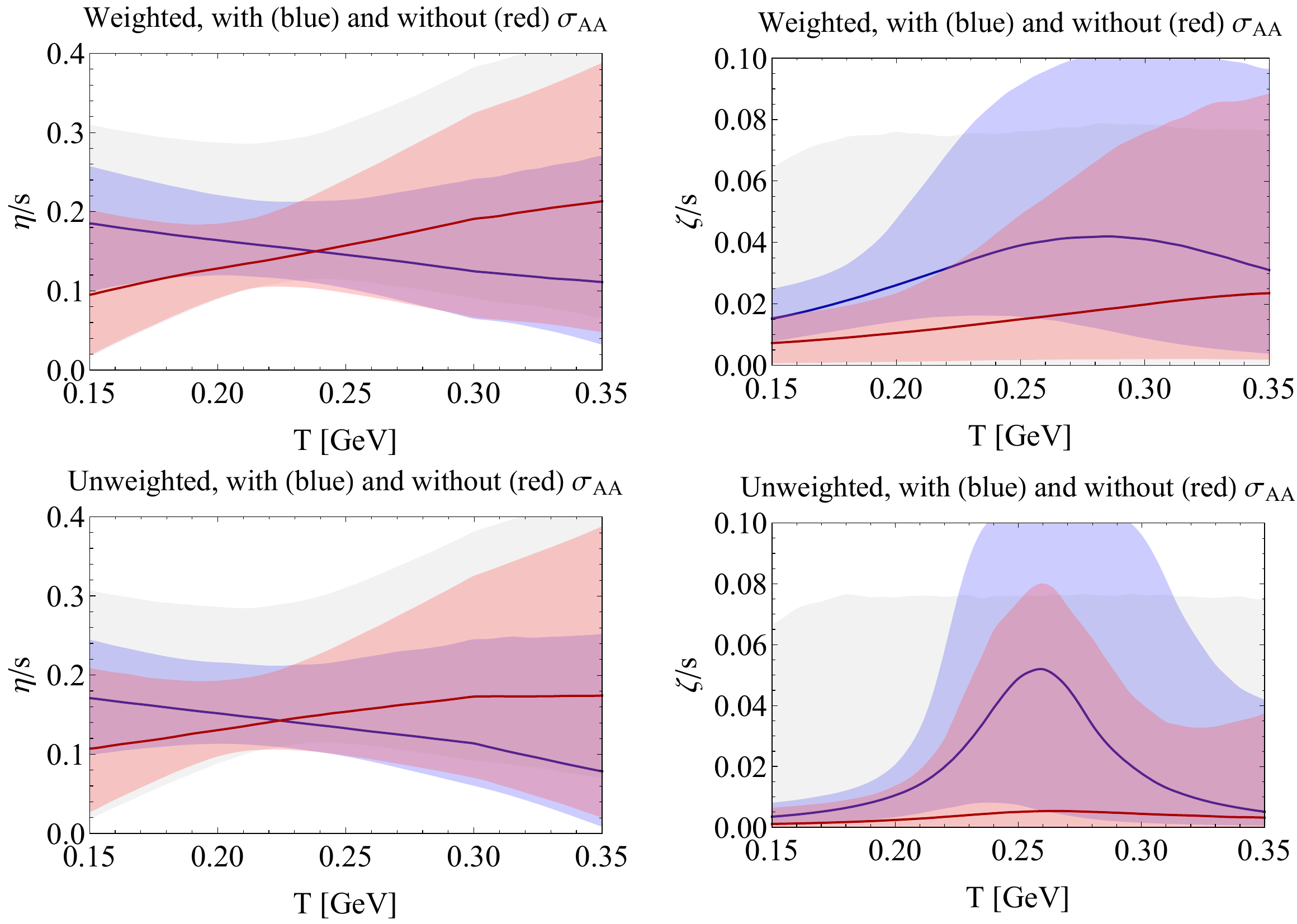}
\caption{\label{fig:viscosities}We show the temperature dependent specific shear and bulk viscosities $\eta/s$ and $\zeta/s$ using the posterior distributions from four different global analyses. This includes (blue) or excludes (red) $\saa$, and includes weighting (top) or does not use weighting (bottom)\@. Unweighted viscosities are more precise (especially for $\zeta/s$), but do not necessarily give the correct nucleon width when considering $\saa{}$.}
\end{figure}

\begin{table}
\begin{tabular}{ccc}
\hline
\hline
& $\sigma_\text{PbPb}$ [b] & $\sigma_{p\text{Pb}}$ [b] \\
\hline
$\sigma_\text{AA}$ \& weights & $8.02\pm 0.19$ & $2.20\pm 0.06$ \\
weights & $8.95\pm 0.36$ & $2.48\pm 0.10$ \\
$\sigma_\text{AA}$ & $8.19\pm 0.19$ & $2.25\pm 0.06$ \\
neither & $8.83\pm 0.29$ & $2.45\pm 0.09$ \\
\hline
ALICE/CMS & $7.67\pm 0.24$ & $2.06\pm 0.08$ \\
\hline
\end{tabular}
\caption{Posterior values for the PbPb and $p$Pb cross sections for the four different fits, compared to the ALICE \cite{2204.10148manual} (PbPb) and CMS \cite{1509.03893} ($p$Pb) values. Theoretical emulation uncertainty is negligible and the uncertainty comes almost entirely from the posterior uncertainty on the nucleon width (see Fig.~\ref{fig:postparams}).\label{tab:cross}}
\end{table}

\begin{figure*}[ht]
\includegraphics[width=0.99\textwidth]{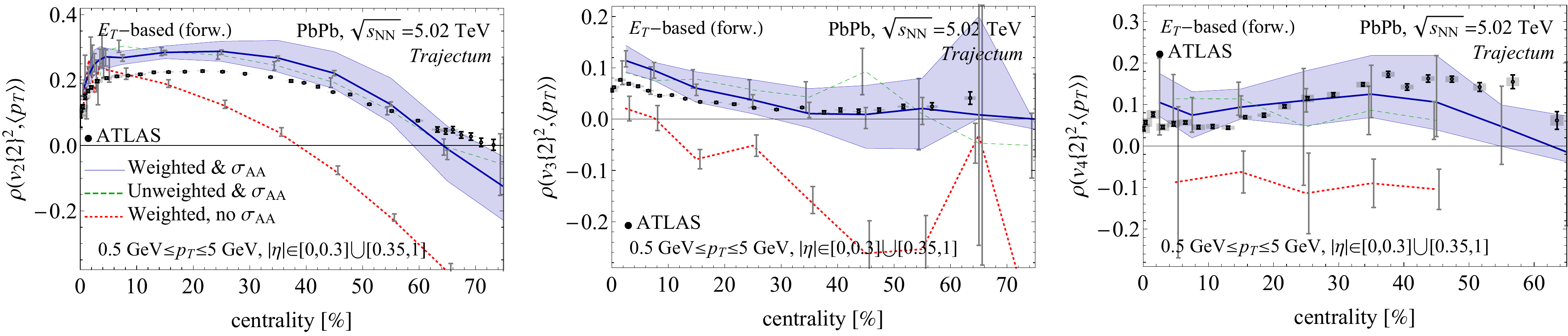}
\caption{\label{fig:rhoobs}We show the correlation between elliptic flow $v_2\{2\}^2$ (left), $v_3\{2\}^2$ (middle) and $v_4\{2\}^2$ (right) with mean transverse momentum $\langle p_T\rangle$ for weighted fit including (solid) and without (dotted) the cross section $\saa{}$ and also the unweighted fit including $\saa{}$ (dashed) for PbPb collisions. We show statistical uncertainties in gray and systematic uncertainties from the posterior as a band (first case only). We also include ATLAS data with systematic (boxes) and statistical uncertainties \cite{2205.00039manual}\@.}
\end{figure*}

Alternatively, we could say that the inconsistency of the nucleon width before and after $\saa{}$ implies that we underestimated the systematic uncertainties. By far the best way would be to update those uncertainties, importantly including all (anti-)correlations between all observables. 
In the current analysis data points from nearby bins are treated as correlated (see also \cite{1804.06469}), although even here there is a certain level of arbitrariness. Moreover, there are theoretical uncertainties from e.g.~particlization that are even in principle hard to quantify.
Inaccurate estimates of correlations can cause the affected class of observables to be given too much or too little relative importance in the Bayesian analysis. The weighting is a way to modify the relative importance of observable classes and hence correct for this effect. This interpretation of extra uncertainty is more applicable to the theoretical uncertainty (where model uncertainties are hard to estimate), but we stress that even experimentally it is often not clear how all systematic uncertainties are correlated among observables.

As mentioned we highly trust $\saa$ on grounds that it is relatively model independent theoretically and in fact only strongly depends on the nucleon width. Weaker but similar arguments can be made for integrated unidentified observables, such as integrated multiplicities, mean transverse momentum and integrated ansiotropic flow coefficients $v_n\{k\}$\@. Particle identified observables are theoretically more difficult to model and here we assign a weight $1/2$\@. We define the weight $\omega$ to mean that we multiply the difference in an observable between theory and experiment by $\omega$. Note that this preserves the correlation matrix. Also transverse momentum ($p_T$) differential observables are more model-dependent, especially at larger $p_T$, and similar arguments can be made for observables in peripheral centrality classes. We hence chose to weight $p_T$-differential observables by an extra factor 1/2 and as well as an extra factor of $(2.5-p_T[\text{GeV}])/1.5$ if $p_T>1\,$GeV. We also multiply the weight for any observable by $(50-c[\%])/50$ if the centrality class $c$ is beyond 50\%.

The posteriors including the weights are shown in Fig.~\ref{fig:postparams} as blue solid (with $\saa$) and red dashed (without $\saa$) curves. Indeed by using lower weights for more model-dependent observables, we see the nucleon width is in agreement with the estimate from $\saa$\@. Also clear is that as expected the lower weights decrease the precision of the posterior parameter determination, which is especially clear for $\tau_\pi$ and $\tau_\text{fs}$\@. As shown in Table~\ref{tab:cross} both including $\saa{}$ and including weights improves the theoretical postdiction of $\saa{}$ so that the agreement is within  1.1 (weighted) or 1.7 (unweighted) standard deviations from the ALICE result for PbPb and 1.5 and 1.9 standard deviations from the CMS result for $p$Pb.

In Fig.~\ref{fig:viscosities} we show the temperature dependent specific shear and bulk viscosities, $\eta/s$ and $\zeta/s$ including (blue) and without $\saa$ (red). Given the smaller nucleon width we expect a larger bulk and shear viscosity to reduce the average radial and elliptic flow that is induced by the larger radial gradient. This is indeed what is observed.

{\bf The nucleon width and observables - }
After showing the $\saa$ updated QGP properties an important question remains. Why do analyses without $\saa$ favor a large nucleon width, and is there an inconsistency given that the nucleon width is small? To answer this question Table~\ref{tab:observables} shows per class of observables the average discrepancy with the experimental result. Naturally, including the $\saa{}$ dramatically improves the agreement with the $\saa{}$ postdiction. What is however perhaps surprising is that the fit of the other observable classes only worsens mildly, on average worsening from 0.83 to 0.89 standard deviations for the weighted case. Virtually all observables get slightly worse, with the notable exception of the mean $p_T$ fluctuations. Naturally, in the weighted case the observables with a lower weight have a worse agreement, which was precisely the motivation for introducing the weights. We note that these deviations do not trivially translate into a $\chi^2$ value, since many observables are highly correlated (see also \cite{1804.06469})\@. A more complete overview of the match of all observables is presented in the Supplemental Material, where indeed by eye it is difficult to see the difference between the fit with and without $\saa{}$\@.

\begin{table}[ht]
\begin{tabular}{cccccc}
\hline
\hline
& $\sigma_\text{AA}$ \& $\omega$ & $\omega$ & $\sigma_\text{AA}$ & neither & $\bar\omega$ \\
\hline
 $dN_\text{ch}/d\eta$ & 0.55 & 0.60 & 1.23 & 1.22 & 1.00 \\
 $dN_{\pi^\pm,k^\pm,p^\pm}/dy$ & 0.76 & 0.70 & 0.60 & 0.57 & 0.48 \\
 $dE_T/d\eta$ & 1.59 & 1.51 & 0.82 & 0.77 & 0.48 \\
 $\langle p_T\rangle_{\text{ch},\pi^\pm,K^\pm,p^\pm}$ & 0.66 & 0.60 & 0.88 & 0.72 & 0.46 \\
 $\delta p_T/\langle p_T\rangle$ & 0.56 & 0.62 & 0.51 & 0.58 & 0.49 \\
 $v_n\{k\}$ & 0.58 & 0.51 & 0.54 & 0.49 & 1.00 \\
 $d^2N_{\pi^\pm}/dy\,dp_T$ & 1.19 & 1.07 & 0.86 & 0.92 & 0.20 \\
 $d^2N_{K^\pm}/dy\,dp_T$ & 1.41 & 1.27 & 0.79 & 0.73 & 0.20 \\
 $d^2N_{p^\pm}/dy\,dp_T$ & 1.35 & 1.21 & 0.73 & 0.67 & 0.25 \\
 $v_2^{\pi^\pm}(p_T)$ & 0.81 & 0.74 & 0.46 & 0.44 & 0.19 \\
 $v_2^{K^\pm}(p_T)$ & 0.92 & 0.89 & 0.55 & 0.55 & 0.19 \\
 $v_2^{p^\pm}(p_T)$ & 0.49 & 0.47 & 0.34 & 0.35 & 0.25 \\
 $v_3^{\pi^\pm}(p_T)$ & 0.65 & 0.57 & 0.69 & 0.57 & 0.24 \\
\hline
 \text{average} & 0.89 & 0.83 & 0.69 & 0.66 & \text{} \\
\hline
$\saa$ & 1.13 & 3.80 & 1.53 & 3.40 & 1.00 \\
\hline
 \end{tabular}
\caption{Average number of standard deviations from experimental data for different classes of observables for the four fits presented in Fig.~\ref{fig:postparams} and the average weight $\bar{\omega}$ per observable class when used in the weighted analysis. Uncertainties include experimental uncertainty and theoretical uncertainty from the emulation (the latter is dominant for the $v_n$ classes). Including the cross section $\saa{}$ in the fit strongly improves the agreement with $\saa{}$ but leads to only a mild worsening for the other observables. \label{tab:observables}}
\end{table}

An excellent test of our model is provided by the triple-differential observable $\rho(v_n\{2\}^2,\langle p_T\rangle)$, which measures the correlation between anisotropic flow and the mean $p_T$ \cite{1601.04513, 2002.08832, 2004.01765}\@. This observable is statistically expensive to compute (we simulate 625k hydro events for 20 parameter settings from the posterior) and hence cannot be included in the Bayesian fit. It moreover sensitively depends on the precise experimental procedure, including cuts on pseudorapidity, cuts on transverse momentum and the method to select centrality bins \cite{2205.00039manual}\@. Nevertheless, the observable is conjectured to be sensitive only to the hydrodynamic initial conditions, and in particular the nucleon width \cite{2111.02908}\@.

Fig.~\ref{fig:rhoobs} presents $\rho(v_n\{2\}^2,\langle p_T\rangle)$ as compared with ATLAS data \cite{2205.00039manual}\@. Due to the expensive nature of this analysis we only include systematic uncertainty from the posterior for the weighted case including $\saa{}$ and show Maximum a Posteriori (MAP) results for the unweighted  with $\saa{}$ and weighed without $\saa{}$ cases. Clearly including $\saa{}$ dramatically improves the description of this observable, which can at least in part be attributed to the smaller nucleon width \cite{2111.02908}\@. A comparison with ALICE data \cite{2111.06106} is included in the Supplemental Material. Curiously the systematic uncertainty for XeXe collisions is significantly larger than for PbPb collisions (see also Supplemental Material).

{\bf Discussion - }
A question deserving further study is in what way our weighting procedure realistically captures the theoretical and experimental uncertainty. Indeed, systematic off-sets on average observables are within one standard deviation from the experimental results and a naive $\chi^2$ would say that the uncertainties are accurate (see also Table \ref{tab:observables} and the Supplemental Material). This, however, ignores the fact that the 653 data points are highly correlated, which is difficult to fully take into account. Also, while observables are on average one standard deviation away from the experimental results, the deviations are not Gaussian. Instead most observables are well within one standard deviation, while a small number deviates significantly. This is a further indication that a naive $\chi^2$ should not be trusted. We hence believe the weighting leads to more physically realistic results, in particular giving a physically realistic $\saa$, nucleon width and bulk viscosity.

On a superficial level the results in this Letter show that the newly measured $\saa$ improves estimates on the nucleon width, and subsequently the transport coefficients. We wish to caution here, however, that this fact should make us rethink the growing popularity of Bayesian analyses. Indeed, prior to this Letter all such analyses ruled out nucleon widths smaller than about 0.8 fm, which in light of this new analysis was not warranted. The question we should ask is if the data really convincingly implied such large nucleon widths? Small changes in many data points (often as small as 0.1 standard deviation) quickly add up to large Bayes factors that can artificially constrain posterior distributions. 

In principle, given accurate estimates of the correlated uncertainties of the data and theoretical model the Bayes posterior is accurate. Note, however, that even underestimating uncertainties by 10\% is enough to cause an otherwise good fit to now be off by about 0.1 standard deviation, which, if this occurs for many data points, will add up, and the Bayesian analysis will try to compensate for this elsewhere. 
In a sense this means that the robustness of the Bayesian posterior depends sensitively on the accuracy of the uncertainty.
We stress that the hardest part is to accurately estimate the theoretical model uncertainty. Currently this only includes emulation uncertainty, which can be sizeable but is not a physical uncertainty. Instead, many modelling choices such as our viscous particlization scheme (see \cite{1804.06469}) or the particulars of the hadronic afterburner have uncertainties that are not included in the analysis. In fact, the full modeling uncertainty is difficult to quantify, though we note that the versatility of our 21-dimensional model attempts to include a wide scope of theoretical uncertainties.

It is perhaps curious that an early Bayesian study found a smaller width of $0.48\pm 0.1$ \cite{1605.03954}\@. As noted in \cite{1804.06469}, however, this is mostly due to having a simplistic initial stage where no radial flow is created during the first $0.4\,$fm$/c$ and due to ignoring bulk viscous corrections during particlization. This earlier value is hence mostly a feature of the model being simplified, albeit in hindsight it got a reasonable nucleon width estimate (but for the wrong reason). 

Another relevant comment not included in this study can be that the nucleon width could depend on their position within the nucleus. It is not so unreasonable that nucleons in the skin of the nucleus could be smaller than nucleons in the center. The size of nucleons in the center would not significantly affect the nucleus-nucleus cross section.

A smaller nucleon width is consistent with other analyses, in particular including the IP-Glasma model \cite{1206.6805, 1603.04349, 1607.01711}\@. Indeed, the IP-Glasma model is based on HERA data, which extracts a transverse two-gluon radius from $J/\psi$ scattering of $0.50\pm 0.03\,$fm  \cite{Caldwell:2010zza}\@. Ref.~\cite{2111.02908} argues that the characteristic nucleon length scale should be of order $0.5\,$fm or smaller, although this can refer to the nucleon width or the constituent width. Since $\saa{}$ is uniquely sensitive to the nucleon width the combination of these measurements leads to a unique characterization of the nucleon spatial profile.

{\bf Acknowledgements - } We thank Somadutta Bhatta, Andrea Dainese, Giuliano Giacalone, Constantin Loizides, Aleksas Mazeliauskas, Rosi Reed, Mike Sas, Urs Wiedemann and You Zhou for interesting discussions.

\bibliographystyle{apsrev4-1}
\bibliography{prl, prlmanual}

\begin{thebibliography}{47}%
\makeatletter
\providecommand \@ifxundefined [1]{%
 \@ifx{#1\undefined}
}%
\providecommand \@ifnum [1]{%
 \ifnum #1\expandafter \@firstoftwo
 \else \expandafter \@secondoftwo
 \fi
}%
\providecommand \@ifx [1]{%
 \ifx #1\expandafter \@firstoftwo
 \else \expandafter \@secondoftwo
 \fi
}%
\providecommand \natexlab [1]{#1}%
\providecommand \enquote  [1]{``#1''}%
\providecommand \bibnamefont  [1]{#1}%
\providecommand \bibfnamefont [1]{#1}%
\providecommand \citenamefont [1]{#1}%
\providecommand \href@noop [0]{\@secondoftwo}%
\providecommand \href [0]{\begingroup \@sanitize@url \@href}%
\providecommand \@href[1]{\@@startlink{#1}\@@href}%
\providecommand \@@href[1]{\endgroup#1\@@endlink}%
\providecommand \@sanitize@url [0]{\catcode `\\12\catcode `\$12\catcode
  `\&12\catcode `\#12\catcode `\^12\catcode `\_12\catcode `\%12\relax}%
\providecommand \@@startlink[1]{}%
\providecommand \@@endlink[0]{}%
\providecommand \url  [0]{\begingroup\@sanitize@url \@url }%
\providecommand \@url [1]{\endgroup\@href {#1}{\urlprefix }}%
\providecommand \urlprefix  [0]{URL }%
\providecommand \Eprint [0]{\href }%
\providecommand \doibase [0]{http://dx.doi.org/}%
\providecommand \selectlanguage [0]{\@gobble}%
\providecommand \bibinfo  [0]{\@secondoftwo}%
\providecommand \bibfield  [0]{\@secondoftwo}%
\providecommand \translation [1]{[#1]}%
\providecommand \BibitemOpen [0]{}%
\providecommand \bibitemStop [0]{}%
\providecommand \bibitemNoStop [0]{.\EOS\space}%
\providecommand \EOS [0]{\spacefactor3000\relax}%
\providecommand \BibitemShut  [1]{\csname bibitem#1\endcsname}%
\let\auto@bib@innerbib\@empty
\bibitem [{\citenamefont {Heinz}\ and\ \citenamefont
  {Snellings}(2013)}]{1301.2826}%
  \BibitemOpen
  \bibfield  {author} {\bibinfo {author} {\bibfnamefont {U.}~\bibnamefont
  {Heinz}}\ and\ \bibinfo {author} {\bibfnamefont {R.}~\bibnamefont
  {Snellings}},\ }\href {\doibase 10.1146/annurev-nucl-102212-170540}
  {\bibfield  {journal} {\bibinfo  {journal} {Ann. Rev. Nucl. Part. Sci.}\
  }\textbf {\bibinfo {volume} {63}},\ \bibinfo {pages} {123} (\bibinfo {year}
  {2013})},\ \Eprint {http://arxiv.org/abs/1301.2826} {arXiv:1301.2826
  [nucl-th]} \BibitemShut {NoStop}%
\bibitem [{\citenamefont {Busza}\ \emph {et~al.}(2018)\citenamefont {Busza},
  \citenamefont {Rajagopal},\ and\ \citenamefont {van~der Schee}}]{1802.04801}%
  \BibitemOpen
  \bibfield  {author} {\bibinfo {author} {\bibfnamefont {W.}~\bibnamefont
  {Busza}}, \bibinfo {author} {\bibfnamefont {K.}~\bibnamefont {Rajagopal}}, \
  and\ \bibinfo {author} {\bibfnamefont {W.}~\bibnamefont {van~der Schee}},\
  }\href {\doibase 10.1146/annurev-nucl-101917-020852} {\bibfield  {journal}
  {\bibinfo  {journal} {Ann. Rev. Nucl. Part. Sci.}\ }\textbf {\bibinfo
  {volume} {68}},\ \bibinfo {pages} {339} (\bibinfo {year} {2018})},\ \Eprint
  {http://arxiv.org/abs/1802.04801} {arXiv:1802.04801 [hep-ph]} \BibitemShut
  {NoStop}%
\bibitem [{\citenamefont {Berges}\ \emph {et~al.}(2021)\citenamefont {Berges},
  \citenamefont {Heller}, \citenamefont {Mazeliauskas},\ and\ \citenamefont
  {Venugopalan}}]{2005.12299}%
  \BibitemOpen
  \bibfield  {author} {\bibinfo {author} {\bibfnamefont {J.}~\bibnamefont
  {Berges}}, \bibinfo {author} {\bibfnamefont {M.~P.}\ \bibnamefont {Heller}},
  \bibinfo {author} {\bibfnamefont {A.}~\bibnamefont {Mazeliauskas}}, \ and\
  \bibinfo {author} {\bibfnamefont {R.}~\bibnamefont {Venugopalan}},\ }\href
  {\doibase 10.1103/RevModPhys.93.035003} {\bibfield  {journal} {\bibinfo
  {journal} {Rev. Mod. Phys.}\ }\textbf {\bibinfo {volume} {93}},\ \bibinfo
  {pages} {035003} (\bibinfo {year} {2021})},\ \Eprint
  {http://arxiv.org/abs/2005.12299} {arXiv:2005.12299 [hep-th]} \BibitemShut
  {NoStop}%
\bibitem [{\citenamefont {Romatschke}\ and\ \citenamefont
  {Romatschke}(2019)}]{1712.05815}%
  \BibitemOpen
  \bibfield  {author} {\bibinfo {author} {\bibfnamefont {P.}~\bibnamefont
  {Romatschke}}\ and\ \bibinfo {author} {\bibfnamefont {U.}~\bibnamefont
  {Romatschke}},\ }\href {\doibase 10.1017/9781108651998} {\emph {\bibinfo
  {title} {{Relativistic Fluid Dynamics In and Out of Equilibrium}}}},\
  Cambridge Monographs on Mathematical Physics\ (\bibinfo  {publisher}
  {Cambridge University Press},\ \bibinfo {year} {2019})\ \Eprint
  {http://arxiv.org/abs/1712.05815} {arXiv:1712.05815 [nucl-th]} \BibitemShut
  {NoStop}%
\bibitem [{\citenamefont {Moreland}\ \emph {et~al.}(2015)\citenamefont
  {Moreland}, \citenamefont {Bernhard},\ and\ \citenamefont
  {Bass}}]{1412.4708}%
  \BibitemOpen
  \bibfield  {author} {\bibinfo {author} {\bibfnamefont {J.~S.}\ \bibnamefont
  {Moreland}}, \bibinfo {author} {\bibfnamefont {J.~E.}\ \bibnamefont
  {Bernhard}}, \ and\ \bibinfo {author} {\bibfnamefont {S.~A.}\ \bibnamefont
  {Bass}},\ }\href {\doibase 10.1103/PhysRevC.92.011901} {\bibfield  {journal}
  {\bibinfo  {journal} {Phys. Rev. C}\ }\textbf {\bibinfo {volume} {92}},\
  \bibinfo {pages} {011901} (\bibinfo {year} {2015})},\ \Eprint
  {http://arxiv.org/abs/1412.4708} {arXiv:1412.4708 [nucl-th]} \BibitemShut
  {NoStop}%
\bibitem [{\citenamefont {Moreland}\ \emph {et~al.}(2020)\citenamefont
  {Moreland}, \citenamefont {Bernhard},\ and\ \citenamefont
  {Bass}}]{1808.02106}%
  \BibitemOpen
  \bibfield  {author} {\bibinfo {author} {\bibfnamefont {J.~S.}\ \bibnamefont
  {Moreland}}, \bibinfo {author} {\bibfnamefont {J.~E.}\ \bibnamefont
  {Bernhard}}, \ and\ \bibinfo {author} {\bibfnamefont {S.~A.}\ \bibnamefont
  {Bass}},\ }\href {\doibase 10.1103/PhysRevC.101.024911} {\bibfield  {journal}
  {\bibinfo  {journal} {Phys. Rev. C}\ }\textbf {\bibinfo {volume} {101}},\
  \bibinfo {pages} {024911} (\bibinfo {year} {2020})},\ \Eprint
  {http://arxiv.org/abs/1808.02106} {arXiv:1808.02106 [nucl-th]} \BibitemShut
  {NoStop}%
\bibitem [{\citenamefont {Bernhard}\ \emph {et~al.}(2019)\citenamefont
  {Bernhard}, \citenamefont {Moreland},\ and\ \citenamefont
  {Bass}}]{Bernhard:2019bmu}%
  \BibitemOpen
  \bibfield  {author} {\bibinfo {author} {\bibfnamefont {J.~E.}\ \bibnamefont
  {Bernhard}}, \bibinfo {author} {\bibfnamefont {J.~S.}\ \bibnamefont
  {Moreland}}, \ and\ \bibinfo {author} {\bibfnamefont {S.~A.}\ \bibnamefont
  {Bass}},\ }\href {\doibase 10.1038/s41567-019-0611-8} {\bibfield  {journal}
  {\bibinfo  {journal} {Nature Phys.}\ }\textbf {\bibinfo {volume} {15}},\
  \bibinfo {pages} {1113} (\bibinfo {year} {2019})}\BibitemShut {NoStop}%
\bibitem [{\citenamefont {Nijs}\ \emph
  {et~al.}(2021{\natexlab{a}})\citenamefont {Nijs}, \citenamefont {van~der
  Schee}, \citenamefont {G\"ursoy},\ and\ \citenamefont
  {Snellings}}]{2010.15130}%
  \BibitemOpen
  \bibfield  {author} {\bibinfo {author} {\bibfnamefont {G.}~\bibnamefont
  {Nijs}}, \bibinfo {author} {\bibfnamefont {W.}~\bibnamefont {van~der Schee}},
  \bibinfo {author} {\bibfnamefont {U.}~\bibnamefont {G\"ursoy}}, \ and\
  \bibinfo {author} {\bibfnamefont {R.}~\bibnamefont {Snellings}},\ }\href
  {\doibase 10.1103/PhysRevLett.126.202301} {\bibfield  {journal} {\bibinfo
  {journal} {Phys. Rev. Lett.}\ }\textbf {\bibinfo {volume} {126}},\ \bibinfo
  {pages} {202301} (\bibinfo {year} {2021}{\natexlab{a}})},\ \Eprint
  {http://arxiv.org/abs/2010.15130} {arXiv:2010.15130 [nucl-th]} \BibitemShut
  {NoStop}%
\bibitem [{\citenamefont {Everett}\ \emph {et~al.}(2021)\citenamefont {Everett}
  \emph {et~al.}}]{2011.01430}%
  \BibitemOpen
  \bibfield  {author} {\bibinfo {author} {\bibfnamefont {D.}~\bibnamefont
  {Everett}} \emph {et~al.} (\bibinfo {collaboration} {JETSCAPE}),\ }\href
  {\doibase 10.1103/PhysRevC.103.054904} {\bibfield  {journal} {\bibinfo
  {journal} {Phys. Rev. C}\ }\textbf {\bibinfo {volume} {103}},\ \bibinfo
  {pages} {054904} (\bibinfo {year} {2021})},\ \Eprint
  {http://arxiv.org/abs/2011.01430} {arXiv:2011.01430 [hep-ph]} \BibitemShut
  {NoStop}%
\bibitem [{\citenamefont {Nijs}\ and\ \citenamefont {van~der
  Schee}(2021)}]{2110.13153}%
  \BibitemOpen
  \bibfield  {author} {\bibinfo {author} {\bibfnamefont {G.}~\bibnamefont
  {Nijs}}\ and\ \bibinfo {author} {\bibfnamefont {W.}~\bibnamefont {van~der
  Schee}},\ }\href@noop {} {\  (\bibinfo {year} {2021})},\ \Eprint
  {http://arxiv.org/abs/2110.13153} {arXiv:2110.13153 [nucl-th]} \BibitemShut
  {NoStop}%
\bibitem [{\citenamefont {Pohl}\ \emph {et~al.}(2010)\citenamefont {Pohl} \emph
  {et~al.}}]{Pohl:2010zza}%
  \BibitemOpen
  \bibfield  {author} {\bibinfo {author} {\bibfnamefont {R.}~\bibnamefont
  {Pohl}} \emph {et~al.},\ }\href {\doibase 10.1038/nature09250} {\bibfield
  {journal} {\bibinfo  {journal} {Nature}\ }\textbf {\bibinfo {volume} {466}},\
  \bibinfo {pages} {213} (\bibinfo {year} {2010})}\BibitemShut {NoStop}%
\bibitem [{\citenamefont {ALICE}(2022)}]{2204.10148manual}%
  \BibitemOpen
  \bibfield  {author} {\bibinfo {author} {\bibnamefont {ALICE}},\ }\href@noop
  {} {\  (\bibinfo {year} {2022})},\ \Eprint {http://arxiv.org/abs/2204.10148}
  {arXiv:2204.10148 [nucl-ex]} \BibitemShut {NoStop}%
\bibitem [{\citenamefont {Alver}\ \emph {et~al.}(2008)\citenamefont {Alver},
  \citenamefont {Baker}, \citenamefont {Loizides},\ and\ \citenamefont
  {Steinberg}}]{0805.4411}%
  \BibitemOpen
  \bibfield  {author} {\bibinfo {author} {\bibfnamefont {B.}~\bibnamefont
  {Alver}}, \bibinfo {author} {\bibfnamefont {M.}~\bibnamefont {Baker}},
  \bibinfo {author} {\bibfnamefont {C.}~\bibnamefont {Loizides}}, \ and\
  \bibinfo {author} {\bibfnamefont {P.}~\bibnamefont {Steinberg}},\ }\href@noop
  {} {\  (\bibinfo {year} {2008})},\ \Eprint {http://arxiv.org/abs/0805.4411}
  {arXiv:0805.4411 [nucl-ex]} \BibitemShut {NoStop}%
\bibitem [{\citenamefont {Loizides}\ \emph {et~al.}(2015)\citenamefont
  {Loizides}, \citenamefont {Nagle},\ and\ \citenamefont
  {Steinberg}}]{1408.2549}%
  \BibitemOpen
  \bibfield  {author} {\bibinfo {author} {\bibfnamefont {C.}~\bibnamefont
  {Loizides}}, \bibinfo {author} {\bibfnamefont {J.}~\bibnamefont {Nagle}}, \
  and\ \bibinfo {author} {\bibfnamefont {P.}~\bibnamefont {Steinberg}},\ }\href
  {\doibase 10.1016/j.softx.2015.05.001} {\bibfield  {journal} {\bibinfo
  {journal} {SoftwareX}\ }\textbf {\bibinfo {volume} {1-2}},\ \bibinfo {pages}
  {13} (\bibinfo {year} {2015})},\ \Eprint {http://arxiv.org/abs/1408.2549}
  {arXiv:1408.2549 [nucl-ex]} \BibitemShut {NoStop}%
\bibitem [{\citenamefont {Loizides}\ \emph {et~al.}(2018)\citenamefont
  {Loizides}, \citenamefont {Kamin},\ and\ \citenamefont
  {d'Enterria}}]{1710.07098}%
  \BibitemOpen
  \bibfield  {author} {\bibinfo {author} {\bibfnamefont {C.}~\bibnamefont
  {Loizides}}, \bibinfo {author} {\bibfnamefont {J.}~\bibnamefont {Kamin}}, \
  and\ \bibinfo {author} {\bibfnamefont {D.}~\bibnamefont {d'Enterria}},\
  }\href {\doibase 10.1103/PhysRevC.97.054910} {\bibfield  {journal} {\bibinfo
  {journal} {Phys. Rev. C}\ }\textbf {\bibinfo {volume} {97}},\ \bibinfo
  {pages} {054910} (\bibinfo {year} {2018})},\ \bibinfo {note} {[Erratum:
  Phys.Rev.C 99, 019901 (2019)]},\ \Eprint {http://arxiv.org/abs/1710.07098}
  {arXiv:1710.07098 [nucl-ex]} \BibitemShut {NoStop}%
\bibitem [{\citenamefont {d'Enterria}\ and\ \citenamefont
  {Loizides}(2021)}]{2011.14909}%
  \BibitemOpen
  \bibfield  {author} {\bibinfo {author} {\bibfnamefont {D.}~\bibnamefont
  {d'Enterria}}\ and\ \bibinfo {author} {\bibfnamefont {C.}~\bibnamefont
  {Loizides}},\ }\href {\doibase 10.1146/annurev-nucl-102419-060007} {\bibfield
   {journal} {\bibinfo  {journal} {Ann. Rev. Nucl. Part. Sci.}\ }\textbf
  {\bibinfo {volume} {71}},\ \bibinfo {pages} {315} (\bibinfo {year} {2021})},\
  \Eprint {http://arxiv.org/abs/2011.14909} {arXiv:2011.14909 [hep-ph]}
  \BibitemShut {NoStop}%
\bibitem [{\citenamefont {Klos}\ \emph {et~al.}(2007)\citenamefont {Klos} \emph
  {et~al.}}]{Klos:2007is}%
  \BibitemOpen
  \bibfield  {author} {\bibinfo {author} {\bibfnamefont {B.}~\bibnamefont
  {Klos}} \emph {et~al.},\ }\href {\doibase 10.1103/PhysRevC.76.014311}
  {\bibfield  {journal} {\bibinfo  {journal} {Phys. Rev. C}\ }\textbf {\bibinfo
  {volume} {76}},\ \bibinfo {pages} {014311} (\bibinfo {year} {2007})},\
  \Eprint {http://arxiv.org/abs/nucl-ex/0702016} {arXiv:nucl-ex/0702016}
  \BibitemShut {NoStop}%
\bibitem [{\citenamefont {Tarbert}\ \emph {et~al.}(2014)\citenamefont {Tarbert}
  \emph {et~al.}}]{1311.0168}%
  \BibitemOpen
  \bibfield  {author} {\bibinfo {author} {\bibfnamefont {C.~M.}\ \bibnamefont
  {Tarbert}} \emph {et~al.},\ }\href {\doibase 10.1103/PhysRevLett.112.242502}
  {\bibfield  {journal} {\bibinfo  {journal} {Phys. Rev. Lett.}\ }\textbf
  {\bibinfo {volume} {112}},\ \bibinfo {pages} {242502} (\bibinfo {year}
  {2014})},\ \Eprint {http://arxiv.org/abs/1311.0168} {arXiv:1311.0168
  [nucl-ex]} \BibitemShut {NoStop}%
\bibitem [{\citenamefont {Skands}\ \emph {et~al.}(2014)\citenamefont {Skands},
  \citenamefont {Carrazza},\ and\ \citenamefont {Rojo}}]{1404.5630}%
  \BibitemOpen
  \bibfield  {author} {\bibinfo {author} {\bibfnamefont {P.}~\bibnamefont
  {Skands}}, \bibinfo {author} {\bibfnamefont {S.}~\bibnamefont {Carrazza}}, \
  and\ \bibinfo {author} {\bibfnamefont {J.}~\bibnamefont {Rojo}},\ }\href
  {\doibase 10.1140/epjc/s10052-014-3024-y} {\bibfield  {journal} {\bibinfo
  {journal} {Eur. Phys. J. C}\ }\textbf {\bibinfo {volume} {74}},\ \bibinfo
  {pages} {3024} (\bibinfo {year} {2014})},\ \Eprint
  {http://arxiv.org/abs/1404.5630} {arXiv:1404.5630 [hep-ph]} \BibitemShut
  {NoStop}%
\bibitem [{\citenamefont {Aamodt}\ \emph {et~al.}(2011)\citenamefont {Aamodt}
  \emph {et~al.}}]{1012.1657}%
  \BibitemOpen
  \bibfield  {author} {\bibinfo {author} {\bibfnamefont {K.}~\bibnamefont
  {Aamodt}} \emph {et~al.} (\bibinfo {collaboration} {ALICE}),\ }\href
  {\doibase 10.1103/PhysRevLett.106.032301} {\bibfield  {journal} {\bibinfo
  {journal} {Phys. Rev. Lett.}\ }\textbf {\bibinfo {volume} {106}},\ \bibinfo
  {pages} {032301} (\bibinfo {year} {2011})},\ \Eprint
  {http://arxiv.org/abs/1012.1657} {arXiv:1012.1657 [nucl-ex]} \BibitemShut
  {NoStop}%
\bibitem [{\citenamefont {Adam}\ \emph
  {et~al.}(2016{\natexlab{a}})\citenamefont {Adam} \emph
  {et~al.}}]{1512.06104}%
  \BibitemOpen
  \bibfield  {author} {\bibinfo {author} {\bibfnamefont {J.}~\bibnamefont
  {Adam}} \emph {et~al.} (\bibinfo {collaboration} {ALICE}),\ }\href {\doibase
  10.1103/PhysRevLett.116.222302} {\bibfield  {journal} {\bibinfo  {journal}
  {Phys. Rev. Lett.}\ }\textbf {\bibinfo {volume} {116}},\ \bibinfo {pages}
  {222302} (\bibinfo {year} {2016}{\natexlab{a}})},\ \Eprint
  {http://arxiv.org/abs/1512.06104} {arXiv:1512.06104 [nucl-ex]} \BibitemShut
  {NoStop}%
\bibitem [{\citenamefont {Adam}\ \emph
  {et~al.}(2016{\natexlab{b}})\citenamefont {Adam} \emph
  {et~al.}}]{1603.04775}%
  \BibitemOpen
  \bibfield  {author} {\bibinfo {author} {\bibfnamefont {J.}~\bibnamefont
  {Adam}} \emph {et~al.} (\bibinfo {collaboration} {ALICE}),\ }\href {\doibase
  10.1103/PhysRevC.94.034903} {\bibfield  {journal} {\bibinfo  {journal} {Phys.
  Rev. C}\ }\textbf {\bibinfo {volume} {94}},\ \bibinfo {pages} {034903}
  (\bibinfo {year} {2016}{\natexlab{b}})},\ \Eprint
  {http://arxiv.org/abs/1603.04775} {arXiv:1603.04775 [nucl-ex]} \BibitemShut
  {NoStop}%
\bibitem [{\citenamefont {Acharya}\ \emph {et~al.}(2019)\citenamefont {Acharya}
  \emph {et~al.}}]{1805.05212}%
  \BibitemOpen
  \bibfield  {author} {\bibinfo {author} {\bibfnamefont {S.}~\bibnamefont
  {Acharya}} \emph {et~al.} (\bibinfo {collaboration} {ALICE}),\ }\href
  {\doibase 10.1016/j.physletb.2019.04.047} {\bibfield  {journal} {\bibinfo
  {journal} {Phys. Lett. B}\ }\textbf {\bibinfo {volume} {793}},\ \bibinfo
  {pages} {420} (\bibinfo {year} {2019})},\ \Eprint
  {http://arxiv.org/abs/1805.05212} {arXiv:1805.05212 [nucl-ex]} \BibitemShut
  {NoStop}%
\bibitem [{\citenamefont {Acharya}\ \emph {et~al.}(2020)\citenamefont {Acharya}
  \emph {et~al.}}]{1910.07678}%
  \BibitemOpen
  \bibfield  {author} {\bibinfo {author} {\bibfnamefont {S.}~\bibnamefont
  {Acharya}} \emph {et~al.} (\bibinfo {collaboration} {ALICE}),\ }\href
  {\doibase 10.1103/PhysRevC.101.044907} {\bibfield  {journal} {\bibinfo
  {journal} {Phys. Rev. C}\ }\textbf {\bibinfo {volume} {101}},\ \bibinfo
  {pages} {044907} (\bibinfo {year} {2020})},\ \Eprint
  {http://arxiv.org/abs/1910.07678} {arXiv:1910.07678 [nucl-ex]} \BibitemShut
  {NoStop}%
\bibitem [{\citenamefont {Abelev}\ \emph {et~al.}(2013)\citenamefont {Abelev}
  \emph {et~al.}}]{1303.0737}%
  \BibitemOpen
  \bibfield  {author} {\bibinfo {author} {\bibfnamefont {B.}~\bibnamefont
  {Abelev}} \emph {et~al.} (\bibinfo {collaboration} {ALICE}),\ }\href
  {\doibase 10.1103/PhysRevC.88.044910} {\bibfield  {journal} {\bibinfo
  {journal} {Phys. Rev. C}\ }\textbf {\bibinfo {volume} {88}},\ \bibinfo
  {pages} {044910} (\bibinfo {year} {2013})},\ \Eprint
  {http://arxiv.org/abs/1303.0737} {arXiv:1303.0737 [hep-ex]} \BibitemShut
  {NoStop}%
\bibitem [{\citenamefont {Abelev}\ \emph
  {et~al.}(2014{\natexlab{a}})\citenamefont {Abelev} \emph
  {et~al.}}]{1407.5530}%
  \BibitemOpen
  \bibfield  {author} {\bibinfo {author} {\bibfnamefont {B.~B.}\ \bibnamefont
  {Abelev}} \emph {et~al.} (\bibinfo {collaboration} {ALICE}),\ }\href
  {\doibase 10.1140/epjc/s10052-014-3077-y} {\bibfield  {journal} {\bibinfo
  {journal} {Eur. Phys. J. C}\ }\textbf {\bibinfo {volume} {74}},\ \bibinfo
  {pages} {3077} (\bibinfo {year} {2014}{\natexlab{a}})},\ \Eprint
  {http://arxiv.org/abs/1407.5530} {arXiv:1407.5530 [nucl-ex]} \BibitemShut
  {NoStop}%
\bibitem [{\citenamefont {Acharya}\ \emph
  {et~al.}(2018{\natexlab{a}})\citenamefont {Acharya} \emph
  {et~al.}}]{1804.02944}%
  \BibitemOpen
  \bibfield  {author} {\bibinfo {author} {\bibfnamefont {S.}~\bibnamefont
  {Acharya}} \emph {et~al.} (\bibinfo {collaboration} {ALICE}),\ }\href
  {\doibase 10.1007/JHEP07(2018)103} {\bibfield  {journal} {\bibinfo  {journal}
  {JHEP}\ }\textbf {\bibinfo {volume} {07}},\ \bibinfo {pages} {103} (\bibinfo
  {year} {2018}{\natexlab{a}})},\ \Eprint {http://arxiv.org/abs/1804.02944}
  {arXiv:1804.02944 [nucl-ex]} \BibitemShut {NoStop}%
\bibitem [{\citenamefont {Adam}\ \emph
  {et~al.}(2016{\natexlab{c}})\citenamefont {Adam} \emph
  {et~al.}}]{1606.06057}%
  \BibitemOpen
  \bibfield  {author} {\bibinfo {author} {\bibfnamefont {J.}~\bibnamefont
  {Adam}} \emph {et~al.} (\bibinfo {collaboration} {ALICE}),\ }\href {\doibase
  10.1007/JHEP09(2016)164} {\bibfield  {journal} {\bibinfo  {journal} {JHEP}\
  }\textbf {\bibinfo {volume} {09}},\ \bibinfo {pages} {164} (\bibinfo {year}
  {2016}{\natexlab{c}})},\ \Eprint {http://arxiv.org/abs/1606.06057}
  {arXiv:1606.06057 [nucl-ex]} \BibitemShut {NoStop}%
\bibitem [{\citenamefont {Acharya}\ \emph
  {et~al.}(2018{\natexlab{b}})\citenamefont {Acharya} \emph
  {et~al.}}]{1805.04390}%
  \BibitemOpen
  \bibfield  {author} {\bibinfo {author} {\bibfnamefont {S.}~\bibnamefont
  {Acharya}} \emph {et~al.} (\bibinfo {collaboration} {ALICE}),\ }\href
  {\doibase 10.1007/JHEP09(2018)006} {\bibfield  {journal} {\bibinfo  {journal}
  {JHEP}\ }\textbf {\bibinfo {volume} {09}},\ \bibinfo {pages} {006} (\bibinfo
  {year} {2018}{\natexlab{b}})},\ \Eprint {http://arxiv.org/abs/1805.04390}
  {arXiv:1805.04390 [nucl-ex]} \BibitemShut {NoStop}%
\bibitem [{\citenamefont {Khachatryan}\ \emph {et~al.}(2016)\citenamefont
  {Khachatryan} \emph {et~al.}}]{1509.03893}%
  \BibitemOpen
  \bibfield  {author} {\bibinfo {author} {\bibfnamefont {V.}~\bibnamefont
  {Khachatryan}} \emph {et~al.} (\bibinfo {collaboration} {CMS}),\ }\href
  {\doibase 10.1016/j.physletb.2016.06.027} {\bibfield  {journal} {\bibinfo
  {journal} {Phys. Lett. B}\ }\textbf {\bibinfo {volume} {759}},\ \bibinfo
  {pages} {641} (\bibinfo {year} {2016})},\ \Eprint
  {http://arxiv.org/abs/1509.03893} {arXiv:1509.03893 [hep-ex]} \BibitemShut
  {NoStop}%
\bibitem [{tra()}]{trajectumcode}%
  \BibitemOpen
  \href@noop {} {}\bibinfo {howpublished}
  {\url{https://sites.google.com/view/govertnijs/trajectum}}\BibitemShut
  {NoStop}%
\bibitem [{\citenamefont {Nijs}\ \emph
  {et~al.}(2021{\natexlab{b}})\citenamefont {Nijs}, \citenamefont {van~der
  Schee}, \citenamefont {G\"ursoy},\ and\ \citenamefont
  {Snellings}}]{2010.15134}%
  \BibitemOpen
  \bibfield  {author} {\bibinfo {author} {\bibfnamefont {G.}~\bibnamefont
  {Nijs}}, \bibinfo {author} {\bibfnamefont {W.}~\bibnamefont {van~der Schee}},
  \bibinfo {author} {\bibfnamefont {U.}~\bibnamefont {G\"ursoy}}, \ and\
  \bibinfo {author} {\bibfnamefont {R.}~\bibnamefont {Snellings}},\ }\href
  {\doibase 10.1103/PhysRevC.103.054909} {\bibfield  {journal} {\bibinfo
  {journal} {Phys. Rev. C}\ }\textbf {\bibinfo {volume} {103}},\ \bibinfo
  {pages} {054909} (\bibinfo {year} {2021}{\natexlab{b}})},\ \Eprint
  {http://arxiv.org/abs/2010.15134} {arXiv:2010.15134 [nucl-th]} \BibitemShut
  {NoStop}%
\bibitem [{\citenamefont {Nijs}\ and\ \citenamefont {van~der
  Schee}()}]{companion}%
  \BibitemOpen
  \bibfield  {author} {\bibinfo {author} {\bibfnamefont {G.}~\bibnamefont
  {Nijs}}\ and\ \bibinfo {author} {\bibfnamefont {W.}~\bibnamefont {van~der
  Schee}},\ }\href@noop {} {\ }\Eprint {http://arxiv.org/abs/220x.xxxxx}
  {arXiv:220x.xxxxx [nucl-th]} \BibitemShut {NoStop}%
\bibitem [{\citenamefont {Abelev}\ \emph
  {et~al.}(2014{\natexlab{b}})\citenamefont {Abelev} \emph
  {et~al.}}]{1405.1849}%
  \BibitemOpen
  \bibfield  {author} {\bibinfo {author} {\bibfnamefont {B.~B.}\ \bibnamefont
  {Abelev}} \emph {et~al.} (\bibinfo {collaboration} {ALICE}),\ }\href
  {\doibase 10.1088/1748-0221/9/11/P11003} {\bibfield  {journal} {\bibinfo
  {journal} {JINST}\ }\textbf {\bibinfo {volume} {9}},\ \bibinfo {pages}
  {P11003} (\bibinfo {year} {2014}{\natexlab{b}})},\ \Eprint
  {http://arxiv.org/abs/1405.1849} {arXiv:1405.1849 [nucl-ex]} \BibitemShut
  {NoStop}%
\bibitem [{\citenamefont {ATLAS}(2022)}]{2205.00039manual}%
  \BibitemOpen
  \bibfield  {author} {\bibinfo {author} {\bibnamefont {ATLAS}},\ }\href@noop
  {} {\  (\bibinfo {year} {2022})},\ \Eprint {http://arxiv.org/abs/2205.00039}
  {arXiv:2205.00039 [nucl-ex]} \BibitemShut {NoStop}%
\bibitem [{\citenamefont {Bernhard}(2018)}]{1804.06469}%
  \BibitemOpen
  \bibfield  {author} {\bibinfo {author} {\bibfnamefont {J.~E.}\ \bibnamefont
  {Bernhard}},\ }\emph {\bibinfo {title} {{Bayesian parameter estimation for
  relativistic heavy-ion collisions}}},\ \href@noop {} {Ph.D. thesis},\
  \bibinfo  {school} {Duke U.} (\bibinfo {year} {2018}),\ \Eprint
  {http://arxiv.org/abs/1804.06469} {arXiv:1804.06469 [nucl-th]} \BibitemShut
  {NoStop}%
\bibitem [{\citenamefont {Bozek}(2016)}]{1601.04513}%
  \BibitemOpen
  \bibfield  {author} {\bibinfo {author} {\bibfnamefont {P.}~\bibnamefont
  {Bozek}},\ }\href {\doibase 10.1103/PhysRevC.93.044908} {\bibfield  {journal}
  {\bibinfo  {journal} {Phys. Rev. C}\ }\textbf {\bibinfo {volume} {93}},\
  \bibinfo {pages} {044908} (\bibinfo {year} {2016})},\ \Eprint
  {http://arxiv.org/abs/1601.04513} {arXiv:1601.04513 [nucl-th]} \BibitemShut
  {NoStop}%
\bibitem [{\citenamefont {Bozek}\ and\ \citenamefont
  {Mehrabpour}(2020)}]{2002.08832}%
  \BibitemOpen
  \bibfield  {author} {\bibinfo {author} {\bibfnamefont {P.}~\bibnamefont
  {Bozek}}\ and\ \bibinfo {author} {\bibfnamefont {H.}~\bibnamefont
  {Mehrabpour}},\ }\href {\doibase 10.1103/PhysRevC.101.064902} {\bibfield
  {journal} {\bibinfo  {journal} {Phys. Rev. C}\ }\textbf {\bibinfo {volume}
  {101}},\ \bibinfo {pages} {064902} (\bibinfo {year} {2020})},\ \Eprint
  {http://arxiv.org/abs/2002.08832} {arXiv:2002.08832 [nucl-th]} \BibitemShut
  {NoStop}%
\bibitem [{\citenamefont {Giacalone}\ \emph {et~al.}(2021)\citenamefont
  {Giacalone}, \citenamefont {Gardim}, \citenamefont {Noronha-Hostler},\ and\
  \citenamefont {Ollitrault}}]{2004.01765}%
  \BibitemOpen
  \bibfield  {author} {\bibinfo {author} {\bibfnamefont {G.}~\bibnamefont
  {Giacalone}}, \bibinfo {author} {\bibfnamefont {F.~G.}\ \bibnamefont
  {Gardim}}, \bibinfo {author} {\bibfnamefont {J.}~\bibnamefont
  {Noronha-Hostler}}, \ and\ \bibinfo {author} {\bibfnamefont {J.-Y.}\
  \bibnamefont {Ollitrault}},\ }\href {\doibase 10.1103/PhysRevC.103.024909}
  {\bibfield  {journal} {\bibinfo  {journal} {Phys. Rev. C}\ }\textbf {\bibinfo
  {volume} {103}},\ \bibinfo {pages} {024909} (\bibinfo {year} {2021})},\
  \Eprint {http://arxiv.org/abs/2004.01765} {arXiv:2004.01765 [nucl-th]}
  \BibitemShut {NoStop}%
\bibitem [{\citenamefont {Giacalone}\ \emph {et~al.}(2022)\citenamefont
  {Giacalone}, \citenamefont {Schenke},\ and\ \citenamefont
  {Shen}}]{2111.02908}%
  \BibitemOpen
  \bibfield  {author} {\bibinfo {author} {\bibfnamefont {G.}~\bibnamefont
  {Giacalone}}, \bibinfo {author} {\bibfnamefont {B.}~\bibnamefont {Schenke}},
  \ and\ \bibinfo {author} {\bibfnamefont {C.}~\bibnamefont {Shen}},\ }\href
  {\doibase 10.1103/PhysRevLett.128.042301} {\bibfield  {journal} {\bibinfo
  {journal} {Phys. Rev. Lett.}\ }\textbf {\bibinfo {volume} {128}},\ \bibinfo
  {pages} {042301} (\bibinfo {year} {2022})},\ \Eprint
  {http://arxiv.org/abs/2111.02908} {arXiv:2111.02908 [nucl-th]} \BibitemShut
  {NoStop}%
\bibitem [{\citenamefont {Acharya}\ \emph {et~al.}(2021)\citenamefont {Acharya}
  \emph {et~al.}}]{2111.06106}%
  \BibitemOpen
  \bibfield  {author} {\bibinfo {author} {\bibfnamefont {S.}~\bibnamefont
  {Acharya}} \emph {et~al.} (\bibinfo {collaboration} {ALICE}),\ }\href@noop {}
  {\  (\bibinfo {year} {2021})},\ \Eprint {http://arxiv.org/abs/2111.06106}
  {arXiv:2111.06106 [nucl-ex]} \BibitemShut {NoStop}%
\bibitem [{\citenamefont {Bernhard}\ \emph {et~al.}(2016)\citenamefont
  {Bernhard}, \citenamefont {Moreland}, \citenamefont {Bass}, \citenamefont
  {Liu},\ and\ \citenamefont {Heinz}}]{1605.03954}%
  \BibitemOpen
  \bibfield  {author} {\bibinfo {author} {\bibfnamefont {J.~E.}\ \bibnamefont
  {Bernhard}}, \bibinfo {author} {\bibfnamefont {J.~S.}\ \bibnamefont
  {Moreland}}, \bibinfo {author} {\bibfnamefont {S.~A.}\ \bibnamefont {Bass}},
  \bibinfo {author} {\bibfnamefont {J.}~\bibnamefont {Liu}}, \ and\ \bibinfo
  {author} {\bibfnamefont {U.}~\bibnamefont {Heinz}},\ }\href {\doibase
  10.1103/PhysRevC.94.024907} {\bibfield  {journal} {\bibinfo  {journal} {Phys.
  Rev. C}\ }\textbf {\bibinfo {volume} {94}},\ \bibinfo {pages} {024907}
  (\bibinfo {year} {2016})},\ \Eprint {http://arxiv.org/abs/1605.03954}
  {arXiv:1605.03954 [nucl-th]} \BibitemShut {NoStop}%
\bibitem [{\citenamefont {Schenke}\ \emph {et~al.}(2012)\citenamefont
  {Schenke}, \citenamefont {Tribedy},\ and\ \citenamefont
  {Venugopalan}}]{1206.6805}%
  \BibitemOpen
  \bibfield  {author} {\bibinfo {author} {\bibfnamefont {B.}~\bibnamefont
  {Schenke}}, \bibinfo {author} {\bibfnamefont {P.}~\bibnamefont {Tribedy}}, \
  and\ \bibinfo {author} {\bibfnamefont {R.}~\bibnamefont {Venugopalan}},\
  }\href {\doibase 10.1103/PhysRevC.86.034908} {\bibfield  {journal} {\bibinfo
  {journal} {Phys. Rev. C}\ }\textbf {\bibinfo {volume} {86}},\ \bibinfo
  {pages} {034908} (\bibinfo {year} {2012})},\ \Eprint
  {http://arxiv.org/abs/1206.6805} {arXiv:1206.6805 [hep-ph]} \BibitemShut
  {NoStop}%
\bibitem [{\citenamefont {M\"antysaari}\ and\ \citenamefont
  {Schenke}(2016{\natexlab{a}})}]{1603.04349}%
  \BibitemOpen
  \bibfield  {author} {\bibinfo {author} {\bibfnamefont {H.}~\bibnamefont
  {M\"antysaari}}\ and\ \bibinfo {author} {\bibfnamefont {B.}~\bibnamefont
  {Schenke}},\ }\href {\doibase 10.1103/PhysRevLett.117.052301} {\bibfield
  {journal} {\bibinfo  {journal} {Phys. Rev. Lett.}\ }\textbf {\bibinfo
  {volume} {117}},\ \bibinfo {pages} {052301} (\bibinfo {year}
  {2016}{\natexlab{a}})},\ \Eprint {http://arxiv.org/abs/1603.04349}
  {arXiv:1603.04349 [hep-ph]} \BibitemShut {NoStop}%
\bibitem [{\citenamefont {M\"antysaari}\ and\ \citenamefont
  {Schenke}(2016{\natexlab{b}})}]{1607.01711}%
  \BibitemOpen
  \bibfield  {author} {\bibinfo {author} {\bibfnamefont {H.}~\bibnamefont
  {M\"antysaari}}\ and\ \bibinfo {author} {\bibfnamefont {B.}~\bibnamefont
  {Schenke}},\ }\href {\doibase 10.1103/PhysRevD.94.034042} {\bibfield
  {journal} {\bibinfo  {journal} {Phys. Rev. D}\ }\textbf {\bibinfo {volume}
  {94}},\ \bibinfo {pages} {034042} (\bibinfo {year} {2016}{\natexlab{b}})},\
  \Eprint {http://arxiv.org/abs/1607.01711} {arXiv:1607.01711 [hep-ph]}
  \BibitemShut {NoStop}%
\bibitem [{\citenamefont {Caldwell}\ and\ \citenamefont
  {Kowalski}(2010)}]{Caldwell:2010zza}%
  \BibitemOpen
  \bibfield  {author} {\bibinfo {author} {\bibfnamefont {A.}~\bibnamefont
  {Caldwell}}\ and\ \bibinfo {author} {\bibfnamefont {H.}~\bibnamefont
  {Kowalski}},\ }\href {\doibase 10.1103/PhysRevC.81.025203} {\bibfield
  {journal} {\bibinfo  {journal} {Phys. Rev. C}\ }\textbf {\bibinfo {volume}
  {81}},\ \bibinfo {pages} {025203} (\bibinfo {year} {2010})}\BibitemShut
  {NoStop}%
\bibitem [{\citenamefont {Aaboud}\ \emph {et~al.}(2020)\citenamefont {Aaboud}
  \emph {et~al.}}]{1904.04808}%
  \BibitemOpen
  \bibfield  {author} {\bibinfo {author} {\bibfnamefont {M.}~\bibnamefont
  {Aaboud}} \emph {et~al.} (\bibinfo {collaboration} {ATLAS}),\ }\href
  {\doibase 10.1007/JHEP01(2020)051} {\bibfield  {journal} {\bibinfo  {journal}
  {JHEP}\ }\textbf {\bibinfo {volume} {01}},\ \bibinfo {pages} {051} (\bibinfo
  {year} {2020})},\ \Eprint {http://arxiv.org/abs/1904.04808} {arXiv:1904.04808
  [nucl-ex]} \BibitemShut {NoStop}%
\end{thebibliography}%

\newpage
\clearpage

\begin{figure*}[ht]
\includegraphics[width=0.8\textwidth]{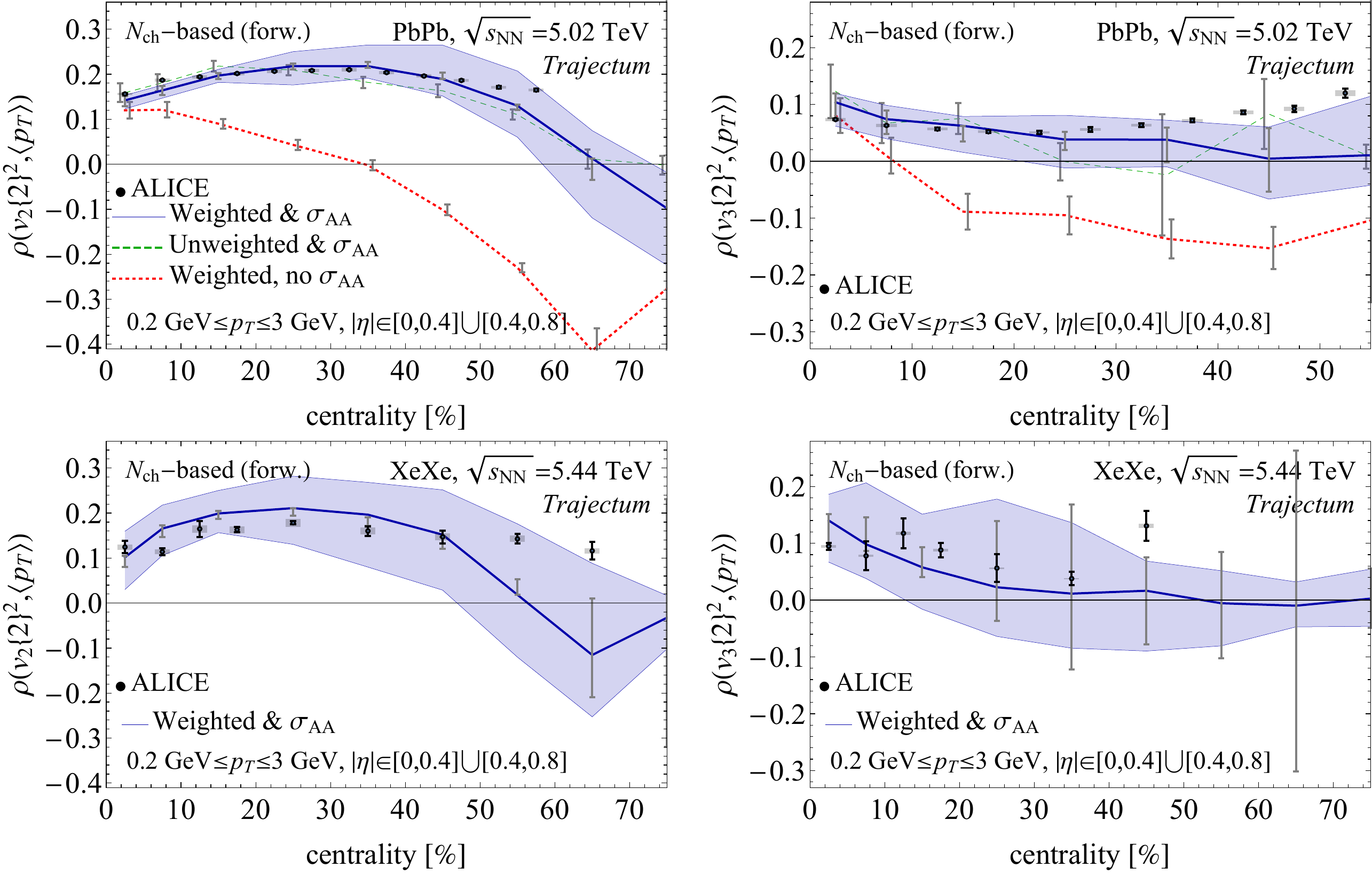}
\caption{\label{fig:rhoobsalice}Similar to Fig.~\ref{fig:rhoobs} we show the comparison with available ALICE data for PbPb (top) and XeXe (bottom) \cite{2111.06106}\@. It is important that ALICE uses different $p_T$ cuts and selects centrality bins based on forward multiplicity instead of forward transverse energy. After taking this into account the weighted and unweighted analyses including $\saa{}$ show satisfactory agreement.}
\end{figure*}

\begin{figure*}[ht]
\includegraphics[width=\textwidth]{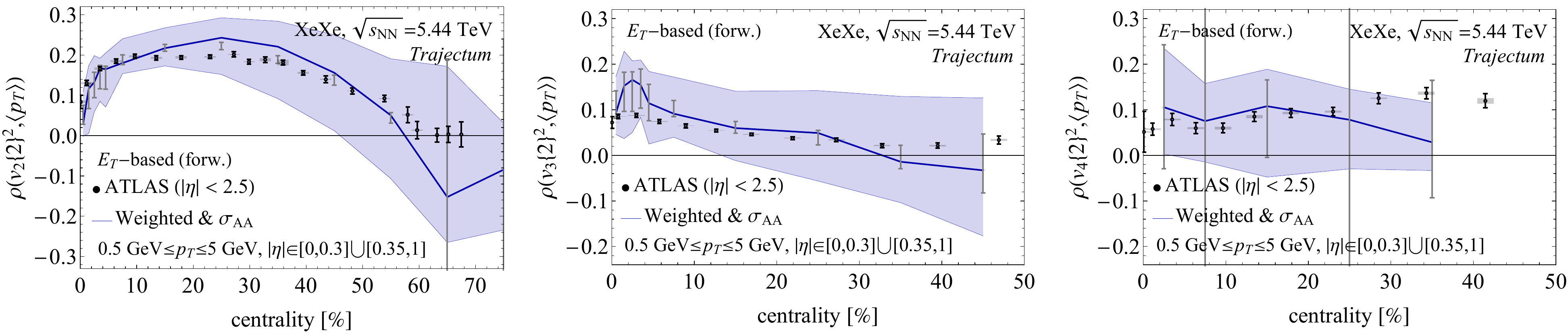}
\caption{\label{fig:rhoobsxe}Similar to Fig.~\ref{fig:rhoobs}, we show the comparison to ATLAS data for XeXe \cite{2205.00039manual}\@. We note that the pseudorapidity cuts used in the theory prediction are different from those used in \cite{2205.00039manual} (see also main text)\@. It is interesting that the systematic uncertainty is significantly larger than for the PbPb case. This is not due to statistical uncertainty (the gray statistical uncertainties are much smaller), but presumably due to the deformed nature of the Xe nucleus.}
\end{figure*}

\section{Supplemental Material}

The $\rho(v_n\{2\}^2,\langle p_T\rangle)$ observables depend sensitively on details such as kinematic cuts on the particles used to compute $\rho$ itself, and in addition to this also on which particles are used for the determination of centrality (offline tracks), and whether these offline tracks are weighted by their transverse energy $E_T$ or not. To illustrate this, in Fig.~\ref{fig:rhoobsalice} (top) we show the results from our analysis for PbPb compared to ALICE data \cite{2111.06106}, which uses different kinematic cuts compared to ATLAS \cite{2205.00039manual}\@. ALICE also uses the forward V0 detectors which measure particle multiplicity, whereas ATLAS uses the forward FCal that measures the total transverse energy $E_T$\@. One can see that the results from both of our fits which include the cross section are in good agreement with the ALICE data, while they are in slight tension with the ATLAS data, especially in the semi-central region (10--30\%)\@.
Here we note that the ALICE and ATLAS data points are largely consistent despite their different experimental cuts (both give $\rho(v_n\{2\}^2,\langle p_T\rangle)$ in the range 0.19--0.23 in this centrality region), whereas for \tr{} we find a significantly larger $\rho(v_n\{2\}^2,\langle p_T\rangle)$ for the ATLAS experimental cuts. 

Indeed, in Fig.~\ref{fig:rhoobs} we find for $\rho(v_2\{2\}^2,\langle p_T\rangle)$ a value in the 10--20\% centrality class of 0.263, whereas for the ALICE comparison we compute only 0.197\@. In \tr{} this difference can be explained by the lower $p_T$ cut (lowering $\rho_2$ by 14\%) and by the $N_{\rm ch}$ versus $E_T$ centrality selection (lowering $\rho_2$ by another 21\%)\@. We note that ATLAS also compared an $N_{\rm ch}$ versus $E_T$ comparison within its own dataset (Fig.~5 in \cite{2205.00039manual}), but here it is essential that $N_{\rm trk}$ is at mid-rapidity and furthermore provides an inferior centrality selector than selecting on $E_T$ \cite{1904.04808}\@. This complicates the comparison, as within $\tr{}$ we find that a mid-rapidity centrality selection increases $\rho_2$ by 13\%, which competes with the decrease due to $N_{\rm ch}$ versus $E_T$ selection such that in the end $\rho_2$ is about 7\% lower for the mid-rapidity centrality selector (somewhat in tension with Fig.~5 of \cite{2205.00039manual}, but see also \cite{1904.04808})\@. Finally, on the other hand, the ATLAS measurement depends sensitively on the $\eta$ cut (Fig.~4 in \cite{2205.00039manual}), by approximately the tension we find in Fig.~\ref{fig:rhoobs}, whereas in $\tr{}$ we do not find any $\eta$-cut dependence in $\rho_2$ (we however present the calculation in Fig.~\ref{fig:rhoobs} for $|\eta|<1$, since this is much cheaper to compute when simulating all 20 parameter settings).

In Fig.~\ref{fig:rhoobsalice} (bottom) and \ref{fig:rhoobsxe}, we also show $\rho(v_n\{2\}^2,\langle p_T\rangle)$ for XeXe collisions at $\sqrt{s_{\rm NN}}=5.44\,$TeV. The normalization of the profile at $5.44\,$TeV is determined by assuming a power-law extrapoliation from the fits at 2.76 and $5.02\,$TeV. We compare results with respectively ALICE \cite{2111.06106} and ATLAS data \cite{2205.00039manual}. The comparison shows good agreement, even though in $\tr{}$ for numerical convenience we use a smaller rapidity cut than ATLAS uses. We verified, however, for a single parameter setting that our boost invariant model does not depend on the rapidity cut. The ATLAS results depend significantly on the rapidity cut and since our MCMC fit is done using mid-rapidity experimental data from ALICE it is preferable to compare with mid-rapidity ATLAS data. For XeXe collisions these are however unfortunately not available and hence the comparison as shown. In the XeXe simulations we use a deformed Woods-Saxon distribution, where the radius $R$ is multiplied by $1 + \beta_2Y_2^0(\theta) + \beta_3Y_3^0(\theta) + \beta_4Y_4^0(\theta)$, with $Y_n^0$ spherical harmonics, $\beta_2 = 0.162$, $\beta_3 = 0$ and $\beta_4 = -0.003$\@. Comparing to the PbPb results, the systematic errors are considerably larger for XeXe, indicating that XeXe is more sensitive to the remaining uncertainty in the posterior parameters.

\begin{figure*}
\includegraphics[width=0.96\textwidth]{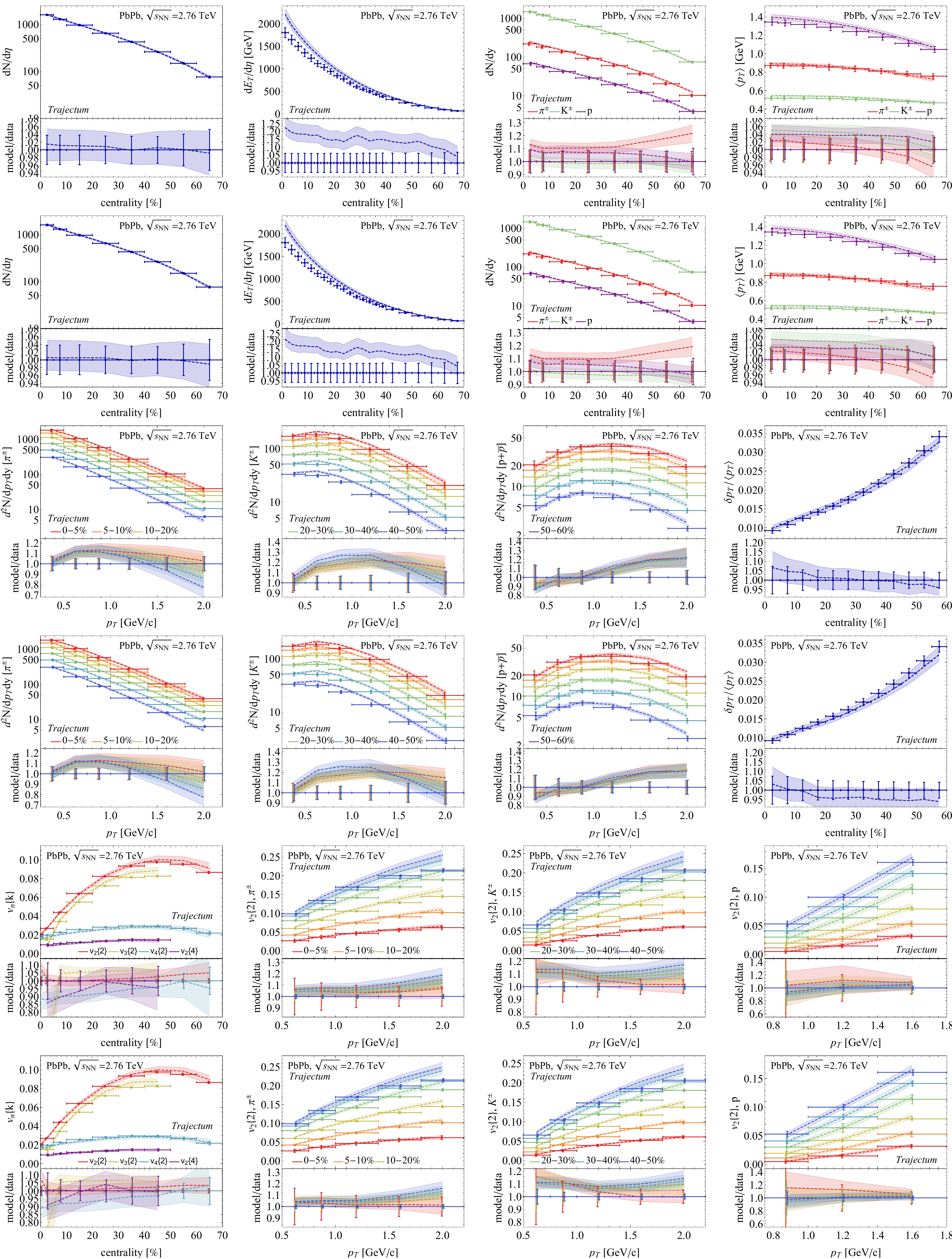}
\caption{\label{fig:posteriorobs1}We show all our 653 data points including a band that is taken from the posterior distribution of our parameters. Odd rows show the results including weights $\saa{}$ (hence having small nucleon width), whereas even rows show the results without $\saa{}$ (also weighted). Often it is hard to notice the difference, but we stress that the odd rows fit $\saa{}$ much better at the cost of a slightly worse fit for the rest of the data points as presented here (continues on next page). }
\end{figure*}

\begin{figure*}
\includegraphics[width=0.96\textwidth]{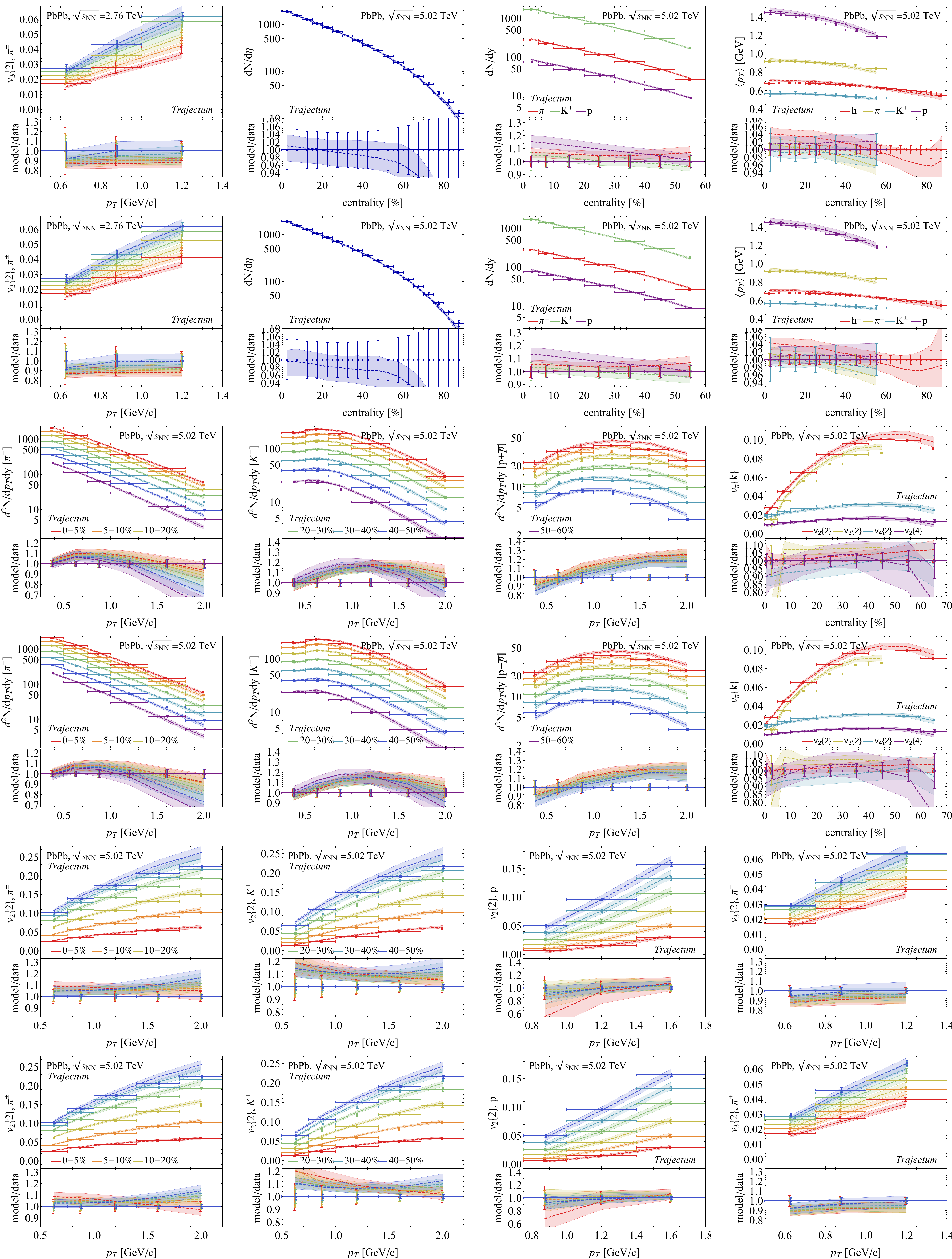}
\caption{\label{fig:posteriorobs2}Continued from previous page.}
\end{figure*}

In Fig.~\ref{fig:posteriorobs1} and \ref{fig:posteriorobs2} we show the experimental data points used together with the posterior distributions from the weighted fit including the new $\saa{}$ (odd rows) and without the $\saa{}$ mesaurement (even rows)\@.
In order of appearance (though 2.76 TeV always appears before 5.02 TeV collisions) we include charged particle multiplicity $dN_\text{ch}/d\eta$ at 2.76 \cite{1012.1657} and 5.02 TeV \cite{1512.06104} and transverse energy $dE_T/d\eta$ at 2.76 TeV \cite{1603.04775}\@. We also include identified yields $dN/dy$ and mean $p_T$ for pions, kaons and protons at 2.76 TeV \cite{1303.0737} and 5.02 TeV \cite{1805.05212,1910.07678}\@. We then include identified transverse momentum spectra for pions, kaons and protons at 2.76 TeV \cite{1303.0737} and 5.02 TeV \cite{1910.07678} in seven coarse grained $p_T$-bins with bin boundaries at $(0.25, 0.5, 0.75, 1.0, 1.4, 1.8, 2.2, 3.0)\,\text{GeV}$\@. Next we show $p_T$ fluctuations, which are only available at 2.76 TeV \cite{1407.5530}\@. Finally we show $p_T$ integrated charged particle anisotropic flow at 2.76 and 5.02 TeV \cite{1804.02944}, as well as the $p_T$-differential identified anisotropic flow coefficient $v_2\{2\}(p_T)$ for pions, kaons and protons, and $v_3\{2\}(p_T)$ for pions at 2.76 TeV \cite{1606.06057} and 5.02 TeV \cite{1805.04390}\@.

\end{document}